\documentclass[prb,showpacs,twocolumn,floatfix]{revtex4}
\usepackage{graphicx}
\begin{document}
\def\deltav{{\mbox{\boldmath{$\delta$}}}}
\draft
\title{Competing Magnetic Phases on a ``Kagom\'e Staircase"}
\author{
G. Lawes,$^1$ M. Kenzelmann,$^{2,3}$ N. Rogado,$^4$ K. H. Kim,$^{1,}$\cite{Kim} G. A. Jorge,$^1$
R. J.  Cava,$^4$ A. Aharony,$^5$ \\ O. Entin-Wohlman,$^5$ A. B. Harris,$^6$
T.  Yildirim,$^3$ Q. Z. Huang,$^3$  S. Park,$^{3,7,}$\cite{Park} C. Broholm,$^{2,3}$ and A. P. Ramirez$^{1,8}$}

\address{$^1$  Los Alamos National Laboratory, Los Alamos, NM 87544}

\address{$^2$ Department of Physics and Astronomy, Johns Hopkins University, Baltimore, MD 21218}

\address{$^3$ NIST Center for Neutron Research, Gaithersburg, MD 20899}

\address{$^4$ Department of Chemistry and Princeton Materials Institute,
Princeton University, Princeton, NJ 08544}

\address{$^5$ School of Physics and Astronomy, Raymond and Beverly Sackler
Faculty of Exact Sciences,Tel Aviv University, Tel Aviv 69978, Israel}

\address{$^6$ Department of Physics and Astronomy,
University of Pennsylvania, Philadelphia, PA, 19104}

\address{$^7$ Department of Materials 
Science and Engineering, University of Maryland, College Park, MD 20742}

\address{$^8$
Bell Labs, Lucent Technologies, 600 Mountain Avenue, Murray Hill, NJ 07974}

\date{\today}

\begin{abstract}
We present thermodynamic and neutron data on Ni$_3$V$_2$O$_8$, a spin-1
system on a kagom\'e staircase.   The extreme degeneracy of the 
kagom\'e antiferromagnet is lifted to produce two incommensurate phases at finite $T$ -- one amplitude modulated, the other helical -- plus a commensurate canted antiferromagnet for $T\rightarrow 0$. The $H-T$ phase diagram is described by a model of competing first and second neighbor interactions with smaller anisotropic terms. $\rm Ni_3V_2O_8$ thus provides an elegant example of order from sub-leading interactions in a highly frustrated system. 
\end{abstract}
\pacs{75.10.Jm, 75.25.+z, 75.30.Kz}
\maketitle

Geometrical magnetic frustration leads to unusual low temperature
spin order and dynamics and presents new challenges for the theoretical
understanding of magnetic systems. Frustrated materials are often characterized
by triangle-based lattices and short-range antiferromagnetic (AF) interactions.\cite{1}
Of particular interest has been magnetism on the two-dimensional (2D) kagom\'e
lattice, which consists of corner-sharing triangles. While the Heisenberg spin-1/2
model appears to have short range spin correlations and a gap to free spinons,\cite{2,3}
the $S\rightarrow \infty$ classical model has N\'eel order with a
$\sqrt 3 \times \sqrt 3$ unit cell at temperature $T=0.\cite{4}$ 
Materials that approximate the kagom\'e AF can be expected to lie close to a
quantum critical point, and indeed early work on the kagom\'e
system SCGO exposed a spin liquid
phase possessing a large fraction (15\%) of the total spin entropy and short range $\sqrt 3 \times \sqrt 3$ order.\cite{5,6} Later work on jarosite systems
showed different ``$q=0$" long range order apparently favored by interlayer
interactions.\cite{7}

\begin{figure}[ht]
\begin{center}
  \includegraphics[width=7.2cm,bbllx=90,bblly=60,bburx=540,bbury=700,angle=0]{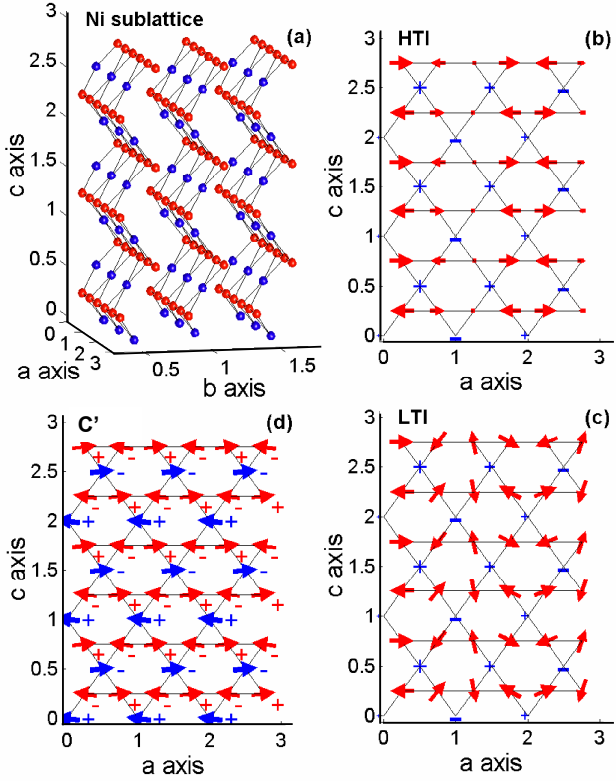}
  \caption{(a) Structure of NVO, showing the cross-tie Ni$_c$ (blue (gray))
and spine Ni$_s$ (red (black)) sites. (b)-(c) indicate the spin structures in the incommensurate phases. + and - indicate spin components along  $\bf b$. Symbol sizes scale with the dipole moment. (d) indicates the symmetry of the low $T$ commensurate spin structure. Spin canting has been exaggerated for clarity and the relative symbol sizes for spine and cross tie spins are not to scale. Subsequent layers are displaced by $({\bf a}+{\bf b})/2$ with spine spins satisfying Eq. (1). Lattice parameters serve as axis length units.   } 
\end{center}
\vspace{-0.25in}
\end{figure}

Here we study Ni$_3$V$_2$O$_8$ (NVO) in which
the $S = 1$ Ni$^{2+}$ spins form the orthorhombic kagom\'e
staircase structure  shown in Fig. 1(a).\cite{structure} This structure has the
coordination and two-dimensionality of the regular kagom\'e lattice,
but the kagom\'e planes are buckled.  The system is particularly
attractive because its complex magnetic phase diagram can be
understood on the basis of an embellished kagom\'{e} spin hamiltonian. The model we introduce also applies to the
isostructural compounds where Ni is replaced by Cu\cite{Cu} or Co.\cite{8}  Although the
symmetry of these compounds is the same as that of NVO, their phase diagrams
are very different.  As indicated below, this difference results
from a small quantitative change in the parameters which dictate how frustration is relieved.

 \begin{figure}[ht]
\begin{center}
  \includegraphics[width=7.4cm,bbllx=105,bblly=60,bburx=465,bbury=550,angle=0]{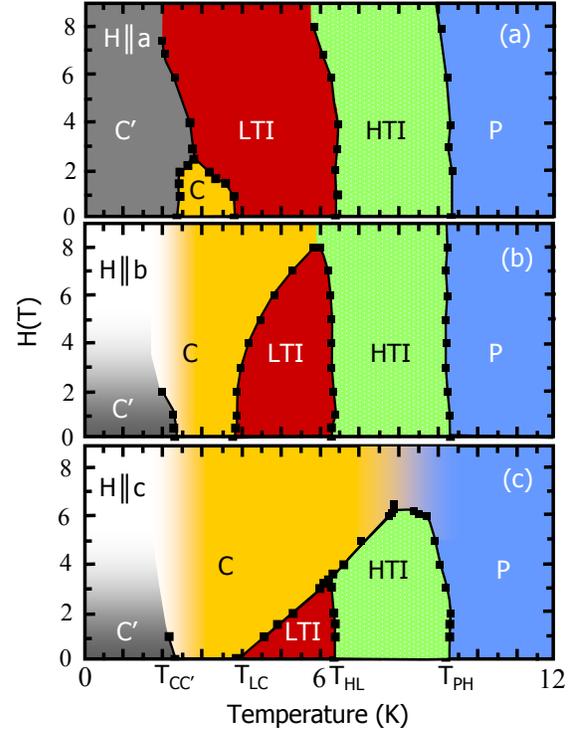} 
  \caption{Phase diagram for NVO as a function of temperature
and magnetic field applied along the three principal
crystallographic directions. For ${\bf H}\parallel{\bf c}$ no true phase boundary separates the P and C phases. White areas were not probed. } 
\end{center}
\vspace{-0.25in}
\end{figure}

A previous study of the magneto-thermal response in polycrystalline
NVO revealed four zero
field phase transitions with $\Theta_W/T_N > 5$, where $\Theta_W$  is the Weiss
constant and $T_N$ the magnetic ordering temperature.\cite{8} In this letter
we report an unexpectedly rich anisotropic field-temperature ($H-T$)
phase diagram (Fig. 2), with high and low temperature incommensurate (IC) phases
(HTI and LTI) and two commensurate (C and C') spin structures.  These magnetic
structures are determined via neutron diffraction.
We also explain the salient features of NVO by a model, in which the spine (Ni$_s$) and cross-tie (Ni$_c$) spins interact
via nearest neighbor (NN) and second  nearest-neighbor (SNN) isotropic
Heisenberg interactions.  In addition, and consistent with crystal symmetry,
it is necessary to take account of the Dzyaloshinskii-Moriya (DM)
interaction and magnetic anisotropy.

Symmetry is key to understanding the ordered phases that spring from the kagom\'e
critical state in NVO.\cite{12}  In the presence of AF ordering on the spine sites, 
isotropic NN interactions produce zero mean field on cross-tie
sites. In this regard, NVO is reminiscent of $\rm Sr_2Cu_3O_4Cl_2$\cite{{10}}  and of
some "ladder" systems of recent interest.\cite{11} However, the structural
anisotropy of NVO induces interactions not usually considered in
frustrated systems. First, because the NiO$_6$ octahedra are
edge-sharing, the NN Ni-O-Ni bond angle is close to 90$^{\rm o}$ so the NN and SNN Ni-Ni
interactions are weak and similar in strength. Second, the symmetry of the crystal structure admits a
DM interaction among the NN spine spins.\cite{10}
Third, anisotropic pseudo dipolar (PD) exchange interactions
between spine and cross-tie spins induce both a uniform and a
staggered moment on the cross-tie sites.\cite{9} These interactions add
to the usual isotropic NN super-exchange interaction to produce
the observed rich $H-T$  phase diagram.

 \begin{figure}[ht]
\begin{center}
  \includegraphics[width=7.5cm,bbllx=30,bblly=100,bburx=390,bbury=525,angle=0]{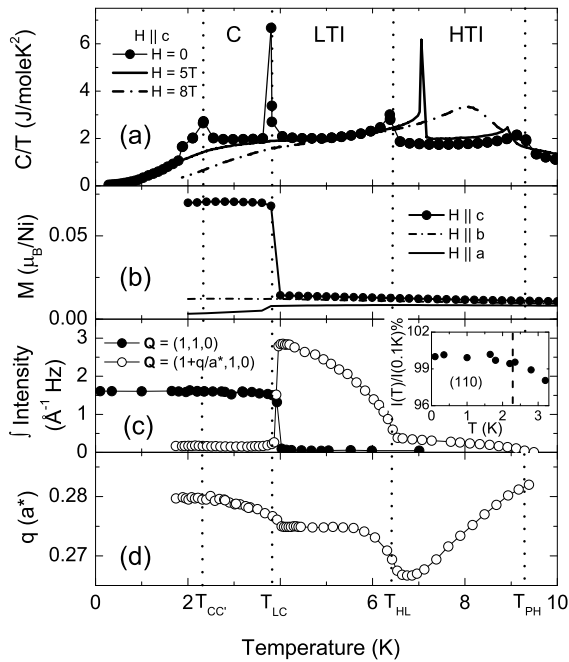}
  \caption{(a) Specific heat of NVO, in zero field and for $H \parallel {\bf c}$.
(b) Longitudinal magnetization versus $T$ for $H=0.1$ T along the three
principal crystallographic directions. (c) Integrated intensity of
commensurate and incommensurate magnetic Bragg peaks at ${\bf Q}=(110)$
and $(1\pm q/a^*,1,0)$, respectively. (d) Temperature dependence of the incommensurate magnetic wave vector. In the C and C' phases we believe the incommensurate peak reflects a meta-stable minority phase as it is only present after cooling through the HTI and LTI phases and can be fully suppressed by field cycling.} 
\end{center}
\vspace{-0.25in}
\end{figure} 

Single crystals of NVO were
grown from a BaO-V$_2$O$_5$ flux and powder samples were synthesized
with standard techniques.\cite{9,8} The uniform magnetization, $M$, was probed using a
commercial SQUID magnetometer. The specific heat, $C$, was measured
with a commercial calorimeter using the relaxation method for $T > 2$
K and the semi-adiabatic method for lower $T$.  Powder and single crystal neutron
diffraction measurements were carried out at the NIST Center for Neutron Research.\cite{9} The space group of NVO is Cmca (No. 64)\cite{structure} with lattice parameters a=5.92197(3) \AA\ b=11.37213(7) \AA, and c=8.22495(5) \AA\ at $T=1.5$ K. Throughout we index wave vectors in the orthorhombic reciprocal lattice with $a^*=2\pi/a$, $b^*=2\pi/b$, and $c^*=2\pi/c$. Representative specific heat data are in Fig. 3(a) for a magnetic field ($H$) of
0, 5, and 8 T along $\bf c$. As in
previous zero field measurements on powder samples, there
are four peaks in $C(T)$.\cite{8}
The entropy reduction associated with these phase transitions is determined by
$\Delta S = \int_0^{50K} (C/T) dT$, after subtracting an estimate of
the lattice contribution obtained from the  non-magnetic structural
analog Zn$_3$V$_2$O$_8$. We find  $\Delta S \approx 7.9$ J/mole K, or
87\% of Rln3, which is close to that expected for ordering among spin-1
Ni$^{2+}$ ions.  We infer that the specific heat peaks mark phase transitions to unique structures involving the Ni$^{2+}$ spin-1 degrees of freedom.
The $H=0$ peaks at  2.2 K,  6.3 K, and  9.1 K indicate second order phase transitions,
whereas the 3.9 K peak marks a first order transition.

Through extensive specific heat measurements,
we determined the phase boundaries shown in Fig. 2. These were
confirmed by susceptibility $(\chi)$ and magnetization measurements (see Fig. 3(b)),
which provide additional clues to the nature of the phases. The
susceptibility exhibits significant magnetic anisotropy.  As $T$ is reduced
and the C phase is
entered, there is a sharp jump in $M$, up to 3.5\% of the Ni$^{2+}$ saturation
moment for  ${\bf H} \parallel {\bf c}$, which indicates a weak ferromagnetic (FM)
moment along $\bf c$.  With ${\bf H} \parallel {\bf a}$,  there is
a sharp drop in $M$. Finally, for ${\bf H} \parallel {\bf b}$,
there is no sharp feature indicating no FM moment
along $\bf b$. A surprising result of this study is that the $T_{PH}$ and
$T_{HL}$ transitions do not produce observable anomalies in $\chi (T)$. In a field of 0.1 T the magnetization anomaly at $T_{PH}$ is less than $4\times 10^{-5}~\mu_B$/Ni or 0.3\% of the signal while it is less than $4\times 10^{-6}~\mu_B$/Ni or 0.03\%
at $T_{HL}$.  Nonlinear susceptibility measurements likewise produced no indication
of these phase transitions.

Neutron diffraction, however, reveals temperature dependent
magnetic Bragg peaks at ${\bf Q}$ = $(2n+1 \pm q/a^*,2m+1,0)$ and ${\bf Q}$=$(2n+1 \pm
q/a^*,2m+1,2m+1)$ for $T_{LC} < T < T_{PH}$.
The peaks are resolution limited indicating a correlation length in excess of 500 \AA .  The $T$-dependence of the
peak intensities is shown in Fig. 3(c). Anomalies are apparent at the three high $T$ transitions,
$T_{PH}$, $T_{HL}$, and $T_{LC}$, and the peaks vanish in the $H= 0$
paramagnetic (P) phase. The absence of an anomaly to the level of 0.5\% (see inset) in the $T$-dependence
of the (110) magnetic Bragg peak through the phase transition at $T_{CC'}$ indicates that this transition involves degrees of
freedom that are decoupled from the prevailing AF order. Fig. 3(b) shows that the weak FM moment is also 
unchanged through this transition. Nonetheless we believe the specific heat anomaly at $T_{CC'}$ is intrinsic as it was observed in all samples studied (1 powder and 5 crystals). 

The $T$-dependence of the characteristic magnetic wave vector, $q$,
is shown in Fig. 3(d). Again there are anomalies at all the upper
transitions but not at $T_{CC'}$. In phases HTI and LTI, $q$ varies
continuously indicative of an IC magnetic structure. The C phase is commensurate though cooling through phases HTI and LTI yields a metastable remnant of the IC modulation.  To determine the spin structures in the HTI, LTI, and C' phases we collected  zero field (ZF) magnetic Bragg intensity data for 170 peaks in the (hk0) and (hkk) planes at $T=7$ K, 5 K, and  for 70 peaks at $T=0.1$ K after ZF cooling. We analyzed the data using group theoretical classification of the possible spin structures.\cite{Izyumov}

In the HTI phase, we limited consideration to magnetic
structures that form a single irreducible representation of the corresponding
space group, because we reject the
possibility of a multicritical point where more than one
irreducible representation simultaneously become critical.\cite{Izyumov} 
Irreducible representation $\Gamma_4$\cite{9} provides an excellent account of the HTI phase with a reliability coefficient R=17\%. The corresponding magnetic structure is illustrated in Fig. 1(b). At $T=7$ K the wavelength of the a-modulated structure is $\lambda_m=2\pi/(a^*-q)=1.37(1)~a$ with an amplitude vector ${\bf m}_s^4=(1.12(4),0.04(9),0.01(8))\ \mu_B$ for spine spins and ${\bf m}_c^4=(0,-0.3(1),0.00(6))\ \mu_B$ for cross-tie spins. There is a phase shift of $0.4(2)\pi$ between the IC waves on the two sublattices.

The LTI phase contains an additional irreducible
representation, $\Gamma_2$, for the $\bf c$ component of spine spins (Fig. 1(c)). For $T=5$ K the spine spin amplitudes are ${\bf m}_s^4=(1.04(8),0.0(1),0.01(6))\ \mu_B$ and ${\bf m}_s^2=(0,-0.01(6),0.76(7)) \mu_B$. The wavevectors for the two components were experimentally indistinguishable and the phase shift of 0.3(1)$\pi$ indicates an elliptical spiral in the $\bf a-c$ plane. The cross-tie amplitude is ${\bf m}_c^4=(0,-0.8(4),0.1(1))\ \mu_B$ with a phase shift of $0.5(2)\pi$ to ${\bf m}_s^4$. There is no detectable cross tie spin component associated with irreducible representation $\Gamma_2$: ${\bf m}_c^2=(0.0(1),0,0)\ \mu_B$. The reliability coefficient was $R=22$\%. 

In the low temperature commensurate C' phase the data are consistent with the spin structure shown in Fig. 1(d) which corresponds to a mixture of representations $\Gamma_1$ and $\Gamma_7$. While the spine sublattice is fully polarized: ${\bf m}_s^1=(0,0.29(8),0)\ \mu_B$ and ${\bf m}_s^7=(2.28(6),0,0.1(4))\ \mu_B$, this is not the case for the cross tie sites where ${\bf m}_c^1=(-0.26(8),0,0)\ \mu_B$ and ${\bf m}_c^7=(0,0.3(1),0.1(8))\ \mu_B$. In this fit the total magnetization along $\bf c$ was fixed to the value of 0.05 $\mu_B$ per Ni atom as determined from bulk magnetization measurements. The  reliability coefficient was $R=11$ \%. Having two active representations in C' indicates that $T_{CC'}$ could mark the admixture upon cooling of $\Gamma_1$. This scenario is, however, difficult to reconcile with the absence of an anomaly in the $T-$dependence of the (110) Bragg intensity (inset to Fig. 3(c)).  

We now turn to a theoretical interpretation of these results.  First,
note that the dominant AF component of the spine magnetization in all $H=0$ phases satisfies
\begin{eqnarray}
{\bf m}({\bf r}) &=& - {\bf m}({\bf r}\pm {1 \over 2} {\bf b})
= - {\bf m}({\bf r} \pm {1 \over 2} {\bf c} + \delta {\bf b}) \ ,
\end{eqnarray}
where $|\delta |=0.26048(6)$\cite{structure} accounts for the kagom\'{e} plane buckling. This indicates AF
interactions between neighboring spines. The spin-structure within spines is controlled by competing NN and SNN isotropic Heisenberg interactions denoted  $J_1$ and $J_2$.
A mean field treatment\cite{13} indicates that for $J_2 > |J_1|/4$ the spine
Hamiltonian is minimized by a mean field spin modulation with
wave vector, $q$, which satisfies $\cos ((a^*-q)a/2) = -J_1/(4J_2)$.
Putting aside the small $T$-dependence of $q$, we
deduce from the experimental value ($q\approx 0.27 a^*$) in the LTI and HTI phases
that $J_1 \approx 2.6 J_2$. In the presence of easy-axis anisotropy
the highest-temperature ordered phase is predicted,\cite{13} in agreement with our
experiments, to be a longitudinally modulated phase in
which the spins are confined to the easy $\bf a$ axis.
If the anisotropy field $H_A$ is not too large ($H_A < H_1$),
then, as the temperature is lowered the longitudinally modulated phase
gives way to one in which an additional transverse modulated component
of spin appears, growing continuously from zero as the LTI phase is
entered. This scenario is also consistent with our diffraction data.
At still lower temperature the diffraction data indicate
the presence of a commensurate AF phase.  According to mean-field
theory, such a transition can occur for sufficiently large anisotropy,
$H_A> H_2$.\cite{13} Our numerical mean field calculations\cite{15} show that for
$J_1/J_2=2.6$ indeed $H_1 > H_2$, so that there is a range of anisotropy field for which
mean field theory predicts the observed sequence of ZF phase transitions. 

We now discuss some of the finer details of these phases.
From the $M$ versus $H$ data, extrapolated to $H=0$, we
find that in the C phase there is a weak ferromagnetic moment.
Structural considerations show that the DM interaction for a single spine
takes the form
\begin{eqnarray}
{\cal H}_{\rm DM} &=& \sum_n [D_c {\bf c} + (-1)^n D_b {\bf b}] \cdot
[{\bf S}(n) \times {\bf S}(n+1)] \ ,
\end{eqnarray} 
where $n$ labels the spins consecutively along the spine.
$D_b$ gives rise to a linear coupling between the staggered
moment of the spine along  ${\bf a}$  and the
weak ferromagnetic moment of the spine along ${\bf c}$.
This weak ferromagnetic moment can induce a ferromagnetic moment
along ${\bf c}$ on the cross-tie spins via isotropic Heisenberg exchange.
In addition, such a moment on the cross-tie spins can also arise via
PD interaction between the staggered moment
on the spines and a uniform moment on the cross-tie spins.
Symmetry also admits a staggered g-tensor along the spines,
the physical origin and consequences of which are similar to DM
interactions.\cite{14}  The weak ferromagnetism explains
the absence of a phase boundary between the P and C phase for
${\bf H} || {\bf c}$.  In the IC phases, these interactions
would give rise to modulated moments along  ${\bf c}$.
The anisotropic interactions we invoke also
generate couplings between the various IC order
parameters, which result in weak $T$-dependence of the
IC wave vector.\cite{15}

Next we discuss the phase boundaries between the C phase and the IC phases. Barring a multicritical point,
these must be first order transitions.  For ${\bf H} || {\bf c}$
the Zeeman energy, $-HM$, of the FM moment (in the C phase) explains why the transition
temperatures $T_{LC}$ and $T_{HC}$ increase linearly with increasing
$H$. For ${\bf H}\perp{\bf c}$ the Zeeman energy does not appear and the phase boundary of the
C phase should be quadratic in $H$
[$T_N(H)=T_N(0)+\alpha H^2$] as it depends on the differences in
the susceptibilities of the phases involved.  In particular, when
${\bf H} || {\bf a}$, the longitudinal susceptibility of the C
phase is small and the coefficient $\alpha$ is negative, disfavoring
the C phase. The other phase boundaries ($T_{PH}$ and $T_{HL}$)
are also expected to be quadratic in $H$ and
the experimental phase diagrams are consistent with this although
for the HTI phase when ${\bf H}$ is along ${\bf b}$, the coefficient $\alpha$
is unusually small.  This fact is linked to the absence of anomalies in $\chi$ at the HTI phase boundaries. 
Both features may be a consequence of a frustrated and
weakly connected spin system where phase transitions occur from a strongly correlated state with short range AF order. 

In summary, we have studied the phase diagram of the spin-1 kagom\'e
staircase $\rm Ni_3V_2O_8$.  We find that although this phase diagram is
quite complicated, it can be understood on the basis
of a rather simple model which reflects the symmetry of the crystal
structure. The experiments and model offer a specific example of how SNN exchange, easy axis anisotropy, and Dzyaloshinskii-Moriya interactions can induce and control complex low temperature phases in a frustrated magnet. 

We acknowledge support from the LDRD program at LANL and the U.S.-Israel
Binational Science Foundation under grant 2000073 for work at TAU,
NIST, UPenn, and JHU.  The NSF supported work at JHU through DMR-0306940, work at Princeton through DMR-0244254, and work at SPINS through DMR-9986442 \& DMR-9704257. KHK is partially supported by KOSEF through CSCMR.

\bibliographystyle{prsty}

\end{document}